\RequirePackage{fix-cm}

\documentclass[twocolumn]{svjour3}          
\smartqed  
\usepackage{ amssymb }
\usepackage{graphicx}
\usepackage{tikz}

\tikzstyle{startstop} = [rectangle, rounded corners, minimum width=2cm, minimum height=3cm,text centered, fill=green!30]
\tikzstyle{blueRect} = [rectangle, rounded corners, minimum width=2cm, minimum height=3cm,text centered, fill=blue!30]
\tikzstyle{process} = [rectangle, rounded corners, minimum width=2cm, minimum height=3cm, text centered, fill=gray!30]
\tikzstyle{arrow} = [thick,->,>=stealth]

\begin{document}

\title{Implementation of ACTS into sPHENIX Track Reconstruction
}


\author{Joseph D. Osborn \and Anthony D. Frawley \and Jin Huang \and Sookhyun Lee \and Hugo Pereira Da Costa \and Michael Peters \and Chris Pinkenburg \and Christof Roland \and Haiwang Yu } 


\institute{Joseph D. Osborn \at
            Oak Ridge National Laboratory \\
            \email{osbornjd@ornl.gov}
            \and
            Anthony D. Frawley \at
             Department of Physics, Florida State University \\
             \email{afrawley@fsu.edu}           
           \and
           Jin Huang \at
              Brookhaven National Laboratory \\
              \email{jhuang@bnl.gov}
            \and
            Sookhyun Lee \at 
            Department of Physics, University of Michigan, \\
            Department of Physics, Iowa State University \\
            \email{sookhyun@umich.edu }
            \and
            Hugo Periera Da Costa \at
            CEA Saclay University \\
            \email{hugo.pereira-da-costa@cea.fr}
            \and 
            Michael Peters \at
            Massachusetts Institute of Technology \\
            \email{mjpeters@mit.edu}
            \and
            Chris Pinkenburg \at
            Brookhaven National Laboratory \\
            \email{pinkenburg@bnl.gov}
            \and
            Christof Roland \at
            Massachusetts Institute of Technology \\
            \email{christof.roland@cern.ch}
            \and
            Haiwang Yu \at
            Brookhaven National Laboratory \\
            \email{hyu@bnl.gov}
}


\maketitle

\begin{abstract}
sPHENIX is a high energy nuclear physics experiment under construction at the Relativistic Heavy Ion Collider at Brookhaven National Laboratory (BNL). The primary physics goals of sPHENIX are to study the quark-gluon-plasma, as well as the partonic structure of protons and nuclei, by measuring jets, their substructure, and heavy flavor hadrons in $p$$+$$p$, $p$+Au, and Au+Au collisions. sPHENIX will collect approximately 300 PB of data over three run periods, to be analyzed using available computing resources at BNL; thus, performing track reconstruction in a timely manner is a challenge due to the high occupancy of heavy ion collision events. The sPHENIX experiment has recently implemented the A Common Tracking Software (ACTS) track reconstruction toolkit with the goal of reconstructing tracks with high efficiency and within a computational budget of 5 seconds per minimum bias event. This paper reports the performance status of ACTS as the default track fitting tool within sPHENIX, including discussion of the first implementation of a time projection chamber geometry within ACTS.
\end{abstract}

\section{Introduction}\label{intro}

The sPHENIX experiment is a next-generation jet and heavy flavor detector being constructed for operation at the Relativstic Heavy Ion Collider (RHIC) at Brookhaven National Laboratory (BNL)~\cite{Adare:2015kwa}. The primary physics goal of sPHENIX is to study strong force interactions by probing the inner workings of the quark-gluon-plasma (QGP) created in heavy nucleus-nucleus collisions, as outlined in the 2015 Nuclear Science Long-Range Plan~\cite{osti_1296778}. sPHENIX will also probe the structure of protons and nuclei in proton-proton and proton-nucleus collisions to study spin-momentum correlations and hadron formation~\cite{sphenixcoldqcd}. To make these measurements, the detector has been designed as a precision jet and heavy-flavor spectrometer. Jets, and their structure, can resolve strong force interactions at different scales when parton flavor is selected due to the difference in mass between heavy and light quarks. Similarly, the measurement of $\Upsilon(1S)$ and its first two excited states allow different screening temperatures of the QGP to be accessed. To achieve these physics goals, precise tracking capabilities are required.

Delivering the desired physics measurements in the environment that will be provided by RHIC will be a substantial challenge. The accelerator will deliver $\sqrt{s_{_{NN}}}=200$ GeV Au+Au collisions at rates of up to 50 kHz, while sPHENIX will trigger on these events at rates of approximately 15 kHz. Because of the electron drift time in the TPC, at 50 kHz the TPC can contain charge deposited by 2 to 3 collisions at any given time. A central Au+Au event can produce approximately 1,000 particles; thus, the occupancy of the detector in any given bunch crossing averages approximately 10\% but can fluctuate up to 25\% in a central event with pile up. These conditions will lead to approximately 300 PB of data collected over the course of a three year running period. These data will be processed on a computing center at BNL of approximately 200,000 CPU nodes, where each node corresponds to 10 HS06. Charged particle or track reconstruction is generally the most computationally expensive portion of data reconstruction at hadron collider experiments; this reconstruction step scales approximately quadratically with the number of charged particles in the event. This necessitates that the tracking be memory efficient and fast so that all data can be processed in a timely manner. To help meet the computational speed requirements for track reconstruction in an environment where per-event detector hit multiplicities are expected to be $\mathcal{O}$(100,000), the sPHENIX Collaboration has implemented the A Common Tracking Software (ACTS) package as the default track reconstruction toolkit.

The ACTS track reconstruction toolkit~\cite{Gessinger:2020nne,actsGithub} is an actively developed open source software package with contributors from several different particle physics collaborations. ACTS is intended to be an experiment-independent set of track reconstruction tools written in modern \texttt{C++} that is customizable and fast. The development was largely motivated by the High-Luminosity Large Hadron Collider (HL-LHC) that will begin data taking in 2027. sPHENIX expects roughly comparable hit occupancies in the heavy ion environments and rates that RHIC will deliver to what is expected in the $p+p$ program at the HL-LHC, for which ACTS was primarily developed; thus, it is a natural candidate for track reconstruction at sPHENIX. In this paper, the ACTS implementation and track reconstruction performance in sPHENIX will be discussed. This includes the first implementation of a TPC geometry in ACTS. Additionally, the current computational and physics performance of the track reconstruction in sPHENIX will be shown, and future directions and improvements that are actively being developed will be discussed.

\begin{figure}[tbh]
	\centering
	\includegraphics[width=1.0\linewidth,clip]{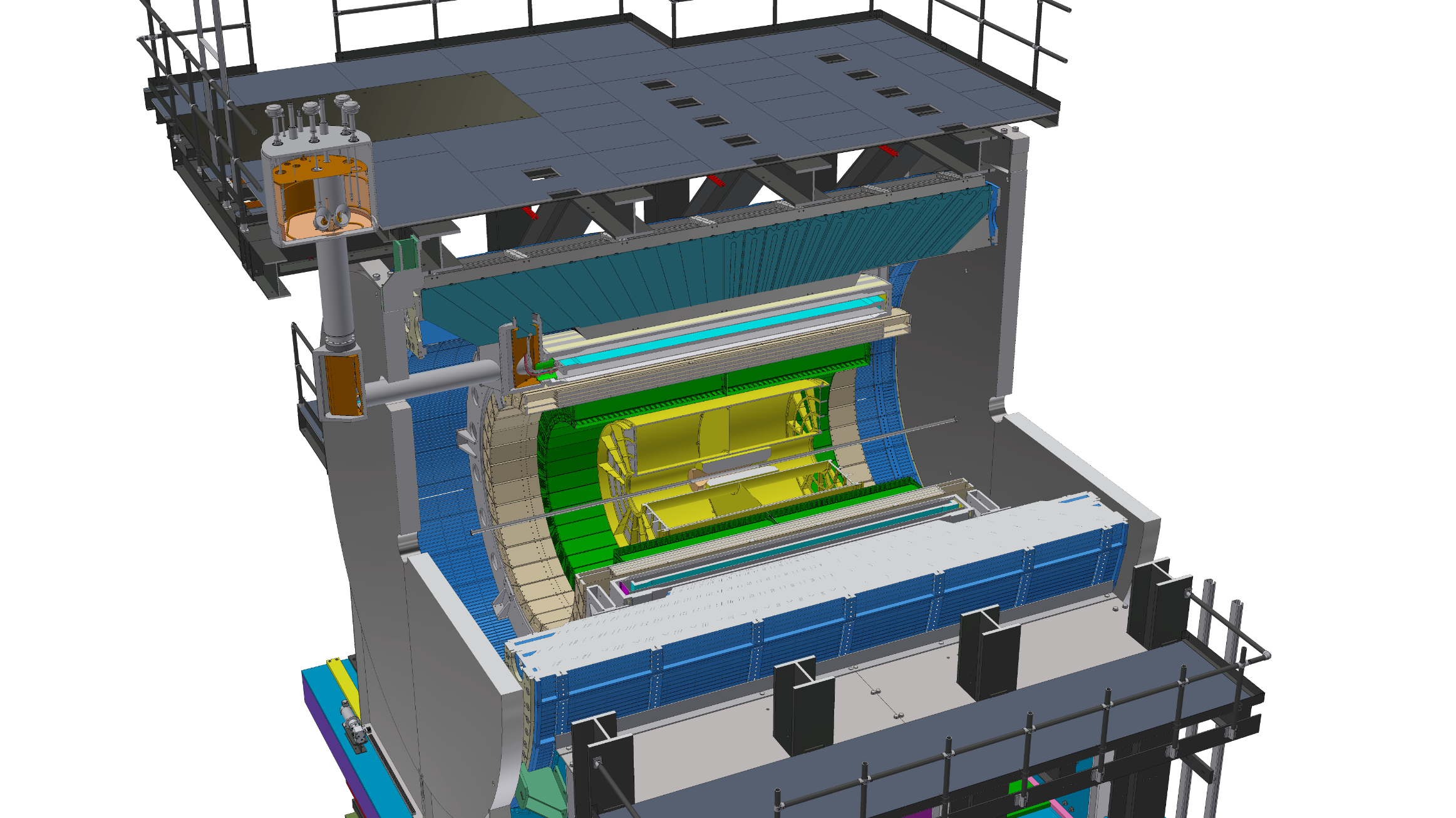}
	\caption{A cutaway engineering diagram of the sPHENIX detector design. The MVTX and INTT are two subdetectors that are composed of silicon staves, shown in orange and grey, respectively. The TPC is a continuous readout GEM-based detector, and the TPC cage is shown in yellow.}
	\label{fig:detector}       
\end{figure}

\begin{figure}[tbh]
	\centering
	\includegraphics[width=0.7\linewidth]{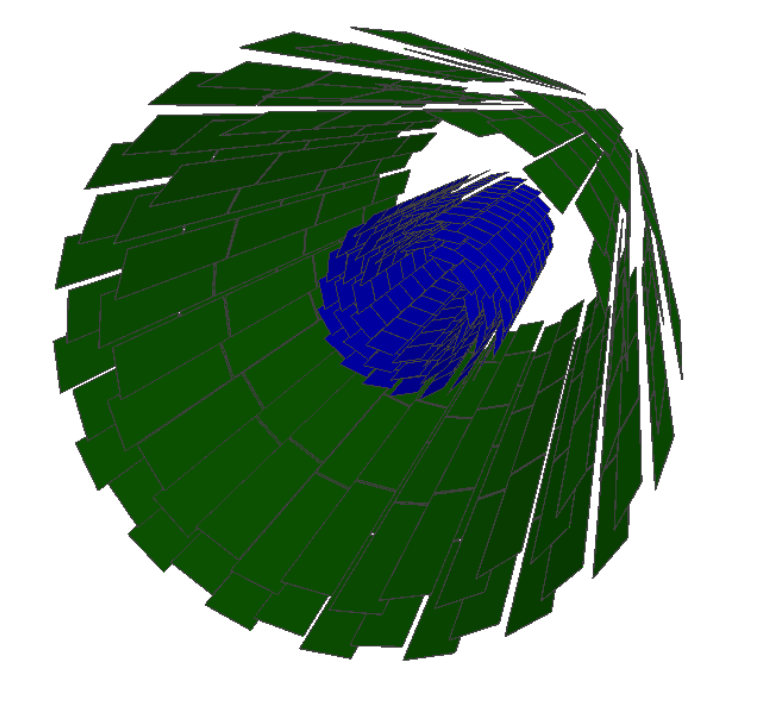}
	\caption{A 3D rendering of the sPHENIX silicon detectors as implemented in ACTS. The small blue layers are the MVTX, and the large green layers are the INTT.}
	\label{fig:actsSiliconDetector}
\end{figure}

\section{sPHENIX Detector and Physics Requirements}\label{detector}

\begin{figure}[tbh]
    \centering
    \hspace*{-2.8cm}\includegraphics[width=1.7\linewidth]{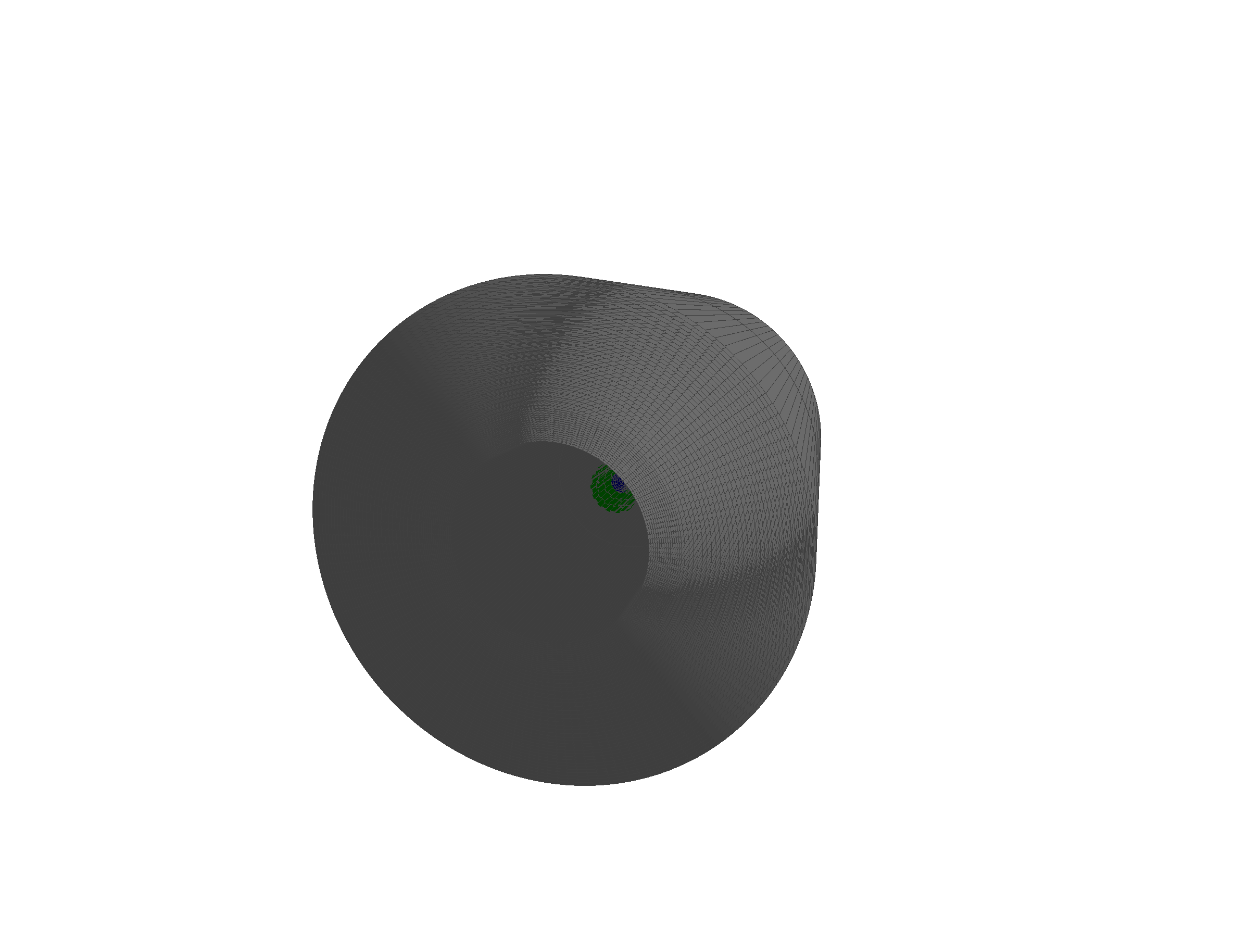}\\
	\includegraphics[width=1.0\linewidth]{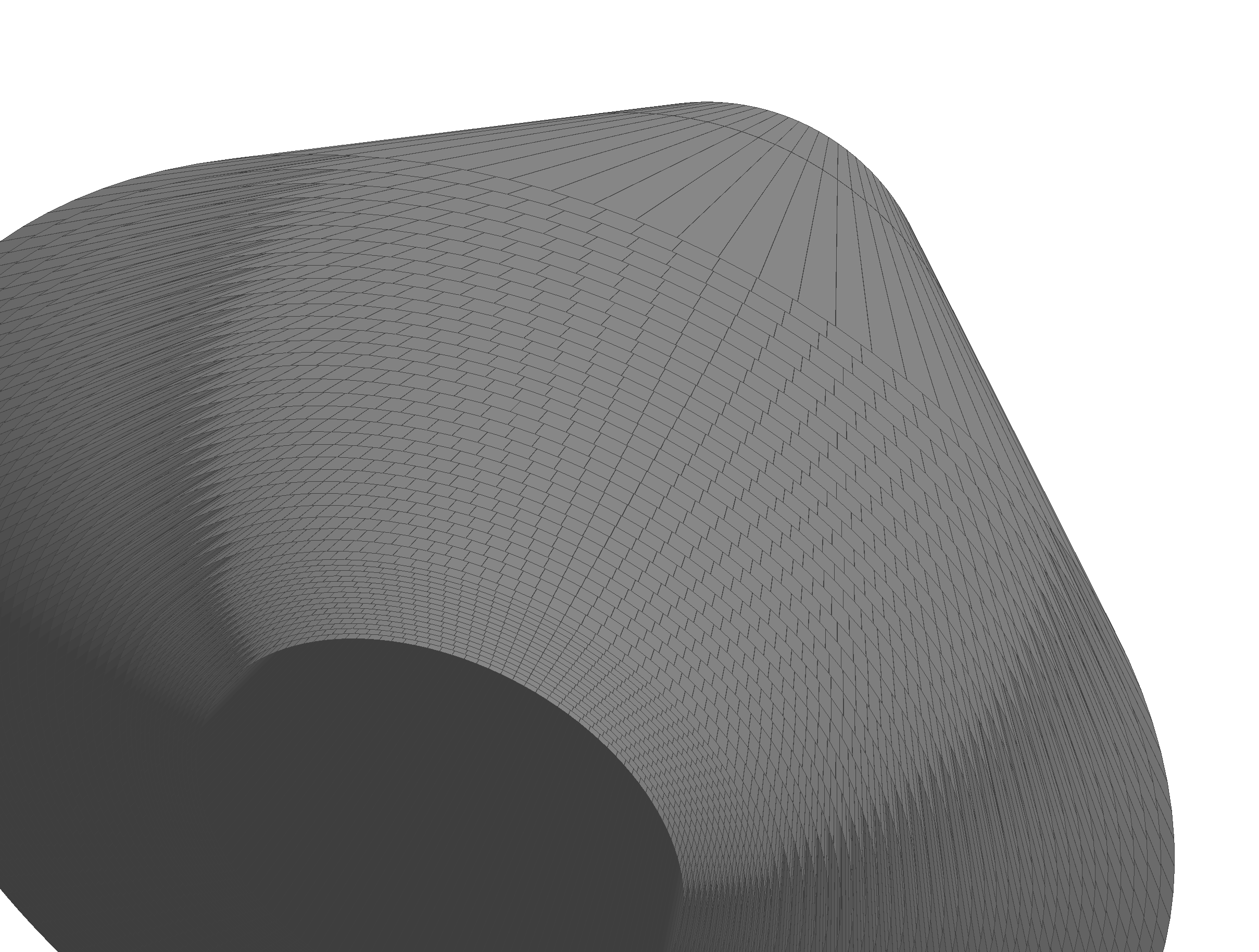}

    \caption{(top) A 3D rendering of the sPHENIX TPC layers as implemented in ACTS. Surfaces are created in place of the TPC pad rows to form cylindrical approximations of the TPC. The MVTX and INTT layers can be seen within the inner TPC layer. (bottom) A zoomed image of the TPC surfaces.}
    \label{fig:actsTpcDetector}
\end{figure}

The sPHENIX spectrometer is a midrapidity barrel detector with full azimuthal and pseudorapidity $|\eta|<1.1$ coverage. The primary subdetectors are three tracking detectors, an electromagnetic calorimeter, and two hadronic calorimeters. An engineering drawing of the detector is shown in Fig.~\ref{fig:detector}. The tracking detectors are a monolithic active pixel sensor (MAPS) based vertex detector (MVTX), a silicon strip detector called the Intermediate Tracker (INTT), and a time projection chamber (TPC). The MVTX has three layers of silicon staves that cover a radial distance of approximately $2<r<4$ cm from the beam pipe. The INTT has two layers of silicon strips and covers approximately $7<r<10$ cm. The TPC is the primary tracking detector within sPHENIX and is a compact, continuous readout gas electron multiplier based TPC. In total, the sPHENIX tracking geometry consists of 53 layers spanning the radial distance from 2$<r<$78 cm. Additional details about each of the detectors can be found in the sPHENIX Technical Design Report~\cite{sphenixtdr}.

The track reconstruction requirements are largely driven by the physics requirements for reconstructing the $\Upsilon(nS)$ states, large transverse momentum jets, and jet substructure. To resolve the first three upsilon states, $e^+e^-$ pairs from upsilon decays must be reconstructed with a mass resolution of less than 100 MeV/$c^2$. Therefore, tracks from upsilon decays must have a resolution of less than $\sim1.2$\%. To resolve high momentum tracks for jet substructure measurements, tracks with $p_T>10$ GeV/$c$ must have a resolution of approximately $\Delta p/p\lesssim 0.2\%\cdot p$ (GeV/$c$). In addition to these requirements, the tracking must be robust against large combinatoric background environments present from pileup, particularly within the TPC. The integration times of the MVTX, INTT, and TPC are approximately $8\mu$s, 100 ns, and 13 $\mu$s, respectively, which provides context for the pileup contributions in each detector when the average RHIC collision rate in a given fill is 50 kHz. Since the MVTX integration time and the TPC drift time are both longer than the bunch spacing provided by RHIC, there is potential for significant out-of-time pileup sampled in the MVTX and TPC.

\section{sPHENIX-ACTS Implementation}\label{ACTSImplementation}

\subsection{Geometry}

The first step for implementing ACTS into the sPHENIX software stack is to properly translate the tracking detector geometry into the analogous ACTS geometry. The main detector element used for track fitting is the \texttt{Acts::Surface}. ACTS has an available ROOT~\cite{ROOT} \texttt{TGeometry} plugin that can take the relevant active \texttt{TGeo} objects and convert them into \texttt{Acts::Surfaces}. Since sPHENIX already has a detailed and well-tested {\sc{Geant 4}}~\cite{Geant4,Geant4_v2} geometry description that uses the ROOT \texttt{TGeoManager}, this plugin was a natural choice. Figure~\ref{fig:actsSiliconDetector} shows the MVTX and INTT active silicon surfaces as implemented within ACTS. The geometry is imported directly from the \texttt{TGeoManager}, so any changes in the sPHENIX {\sc{Geant 4}} description are automatically propagated to the \texttt{Acts::Surface} description.

The sPHENIX TPC geometry is implemented in a different way from the silicon detectors due to the requirement within ACTS that measurements and track states must be associated to a detector surface. This is not ideal for TPC or drift chamber geometries, which utilize a three dimensional volume structure. In the sPHENIX TPC, measurements are readout on the pad planes; however, their truth position can be anywhere within the TPC volume. For this reason, a different approach was taken for the TPC geometry implementation within ACTS. Rather than importing the TPC geometry from the \texttt{TGeoManager}, the ACTS TPC surfaces are constructed individually as plane surfaces that approximate cylinders, as shown in the bottom panel of Fig.~\ref{fig:actsTpcDetector}. For context of the scale of these renderings, the entire length along the $z$ direction of the TPC is 210 cm. The plane surfaces span 3$^\circ$ in azimuth and half the length of the TPC in $z$ and are used to approximate the TPC readout geometry. These surface dimensions were chosen as an optimization of limiting the memory needed for the raw number of surfaces while also maintaining the $r\phi$ cluster resolution of the TPC. Measurements are then associated to these surfaces based on what pad plane they were read out on and where they were physically measured on that pad plane.

\subsection{Track Reconstruction Strategy}

\begin{figure*}[tbh]
    \centering
    \begin{tikzpicture}[node distance=2cm]
    	\node (sphenix) [blueRect] {sPHENIX Object};
	    \node (acts) [process, right of=sphenix, xshift=3cm] {sPHENIX-ACTS Module};
    	\node (actstool) [startstop, right of=acts, xshift=3cm] {ACTS Tool};
    	\begin{scope}[transform canvas={yshift=0.7cm}]
    		\draw [arrow] (sphenix) -- node[anchor=south] {ACTS info} (acts);
    	\end{scope}
    	\begin{scope}[transform canvas={yshift=-0.7cm}]
    		\draw [arrow] (acts) -- node[anchor=north] {Update} (sphenix);
    	\end{scope}
    	\begin{scope}[transform canvas={yshift=0.7cm}]
    		\draw [arrow] (acts) -- node[anchor=south] {Call tool} (actstool);
    	\end{scope}
    	\begin{scope}[transform canvas={yshift=-0.7cm}]
    		\draw [arrow] (actstool) -- node[anchor=north] {ACTS result} (acts);
    	\end{scope}
    \end{tikzpicture}
    \caption{A flow chart demonstrating the sPHENIX-ACTS implementation. Objects within the sPHENIX framework carry raw measurement information, such as the two-dimensional local position of the measured cluster. An sPHENIX-ACTS module serves as a wrapper that interfaces with the ACTS tool, converting and updating the relevant sPHENIX object.}
    \label{fig:actsImplementation}
\end{figure*}
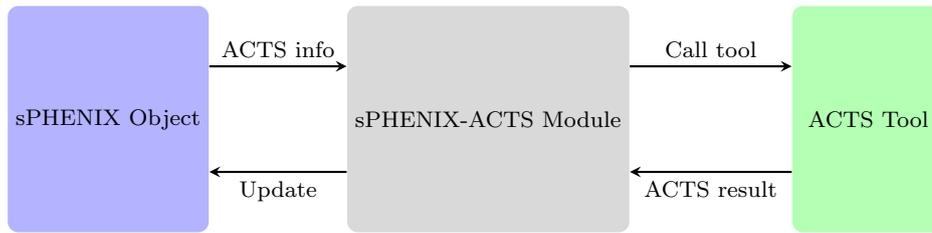

\begin{figure}[tbh]
    \centering
    \includegraphics[width=\linewidth]{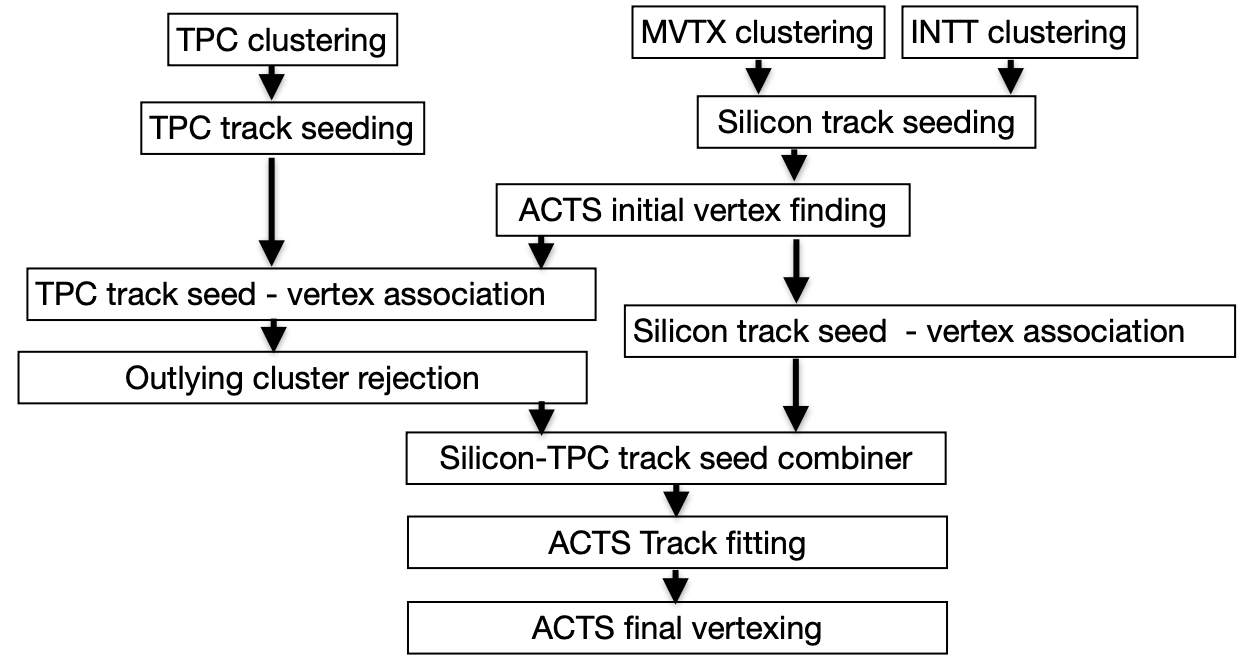}
    \caption{The workflow for track reconstruction in sPHENIX is shown. The workflow flows from top to bottom, starting with clustering in each subsystem and finishing with reconstructed tracks and vertices.}
    \label{fig:tracking_workflow}
\end{figure}

\begin{figure}[tbh]
	\centering
	\includegraphics[width=1\linewidth,clip]{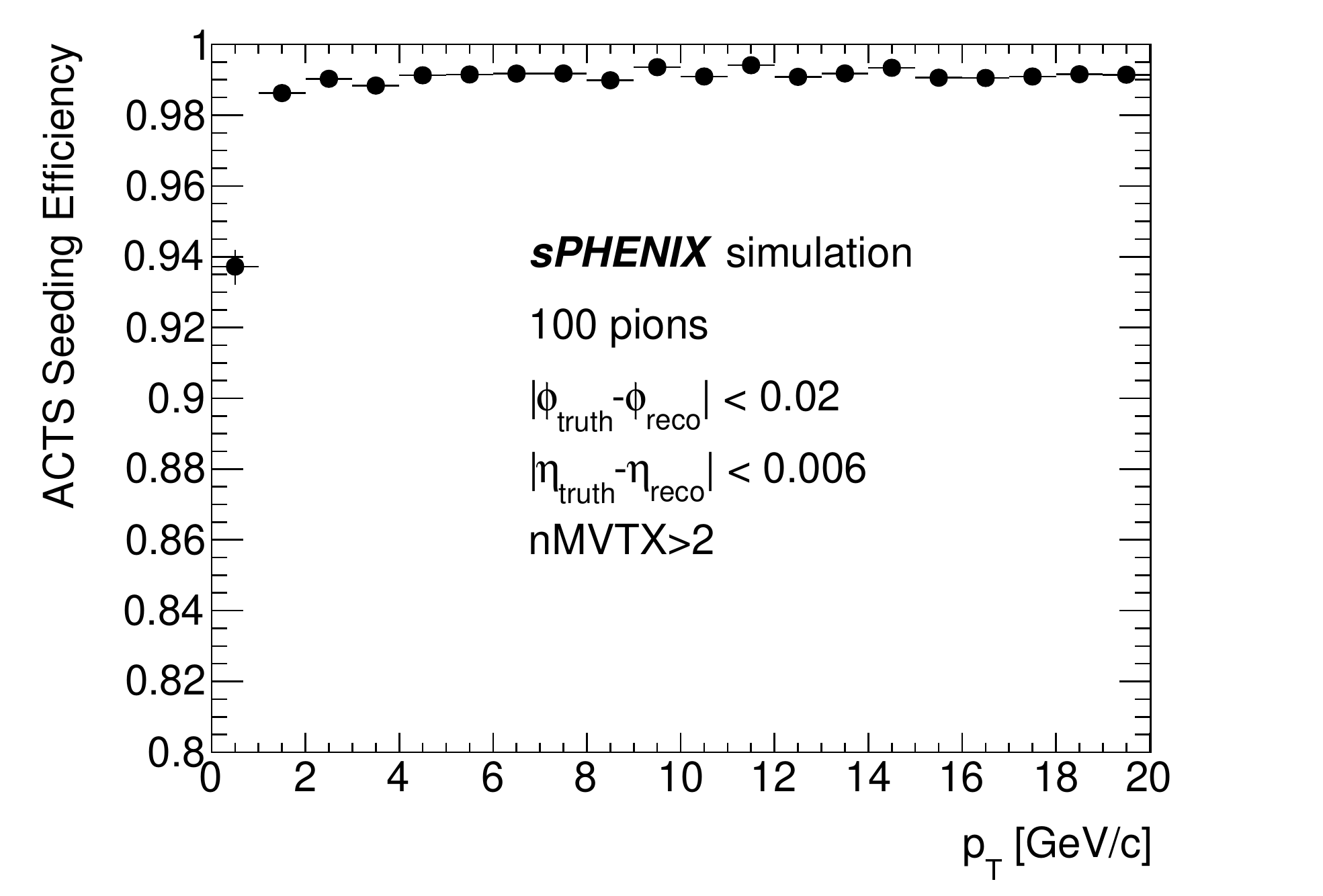}
	\caption{The ACTS seeding efficiency as implemented in the sPHENIX MVTX. The efficiency is defined in the text.}
	\label{fig:seedEff}       
\end{figure}

To perform track reconstruction, the relevant sPHENIX objects are mapped to the corresponding ACTS objects and passed to the ACTS track reconstruction tools. Figure~\ref{fig:actsImplementation} shows a flow chart that demonstrates the software implementation. In practice, the only objects that ACTS requires are the detector geometry and corresponding measurements. Thus, the next step after building the ACTS geometry, as discussed previously, is to translate the sPHENIX measurement objects into ACTS measurement objects associated to the relevant surface. A module that acts as an interface between sPHENIX and the relevant ACTS tool is used to translate the needed information between ACTS and sPHENIX. A variety of ACTS track reconstruction tools exist within the sPHENIX framework that can subsequently be run with the ACTS translated measurements and geometry object~\cite{sphenixgit}.

 \begin{figure}
     \centering
     \includegraphics[width=1.0\linewidth]{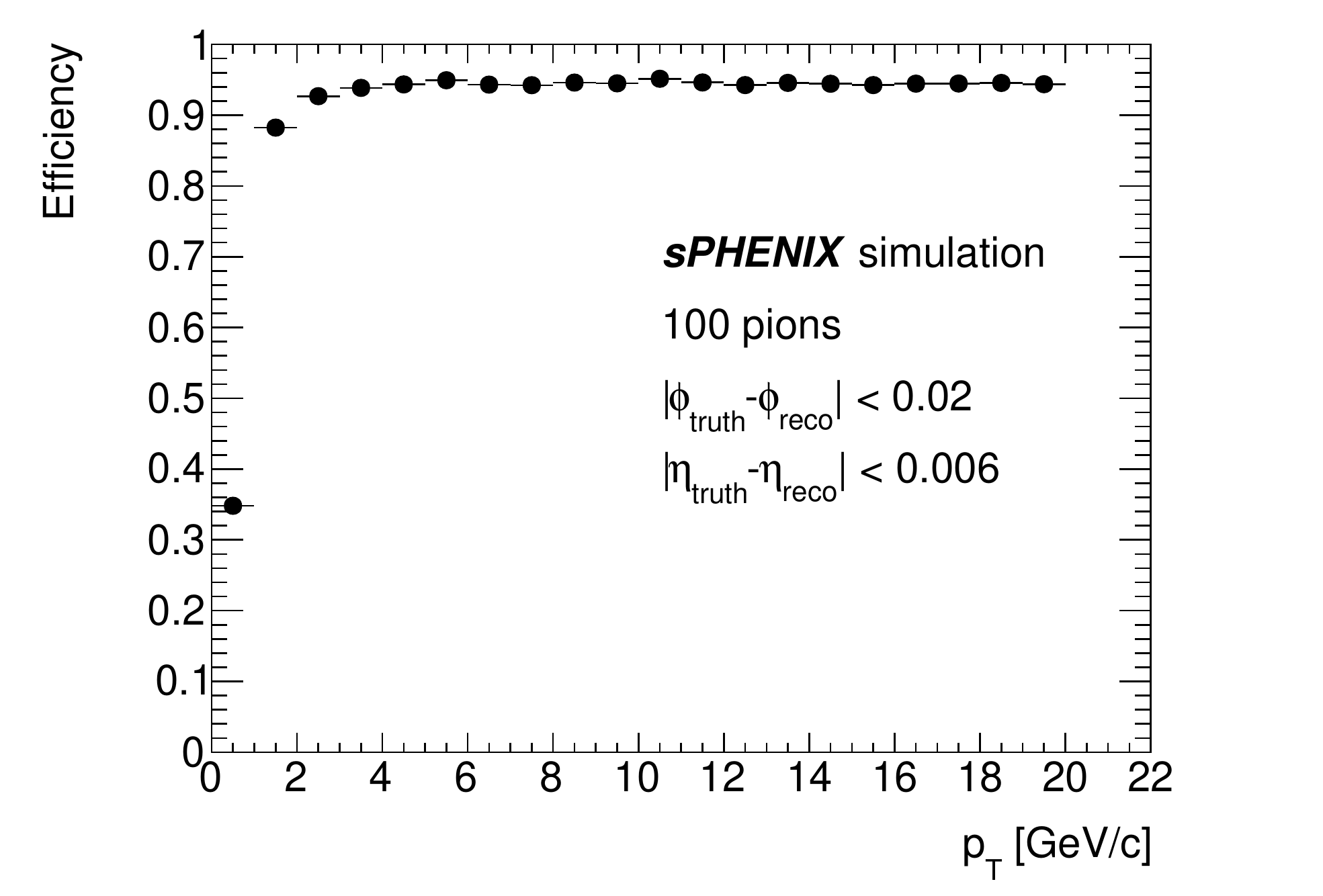}
     \caption{The efficiency of the cellular automaton seeding algorithm as implemented in the sPHENIX TPC. The efficiency is defined in the text.}
     \label{fig:caseeder}
 \end{figure}

 \begin{figure*}[tbh]
 	\centering
 	\includegraphics[width=0.494\linewidth,clip]{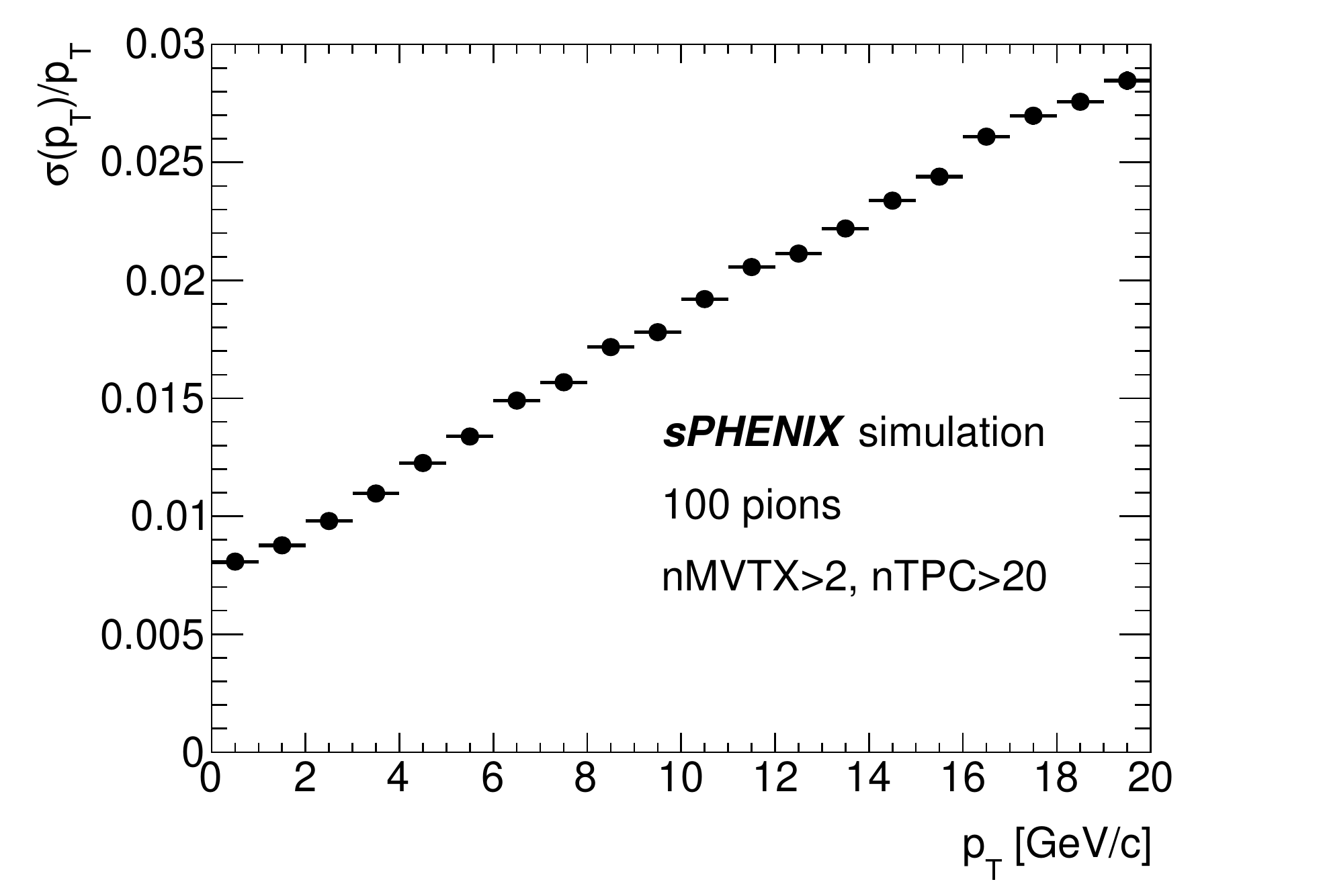}
 	\includegraphics[width=0.494\linewidth,clip]{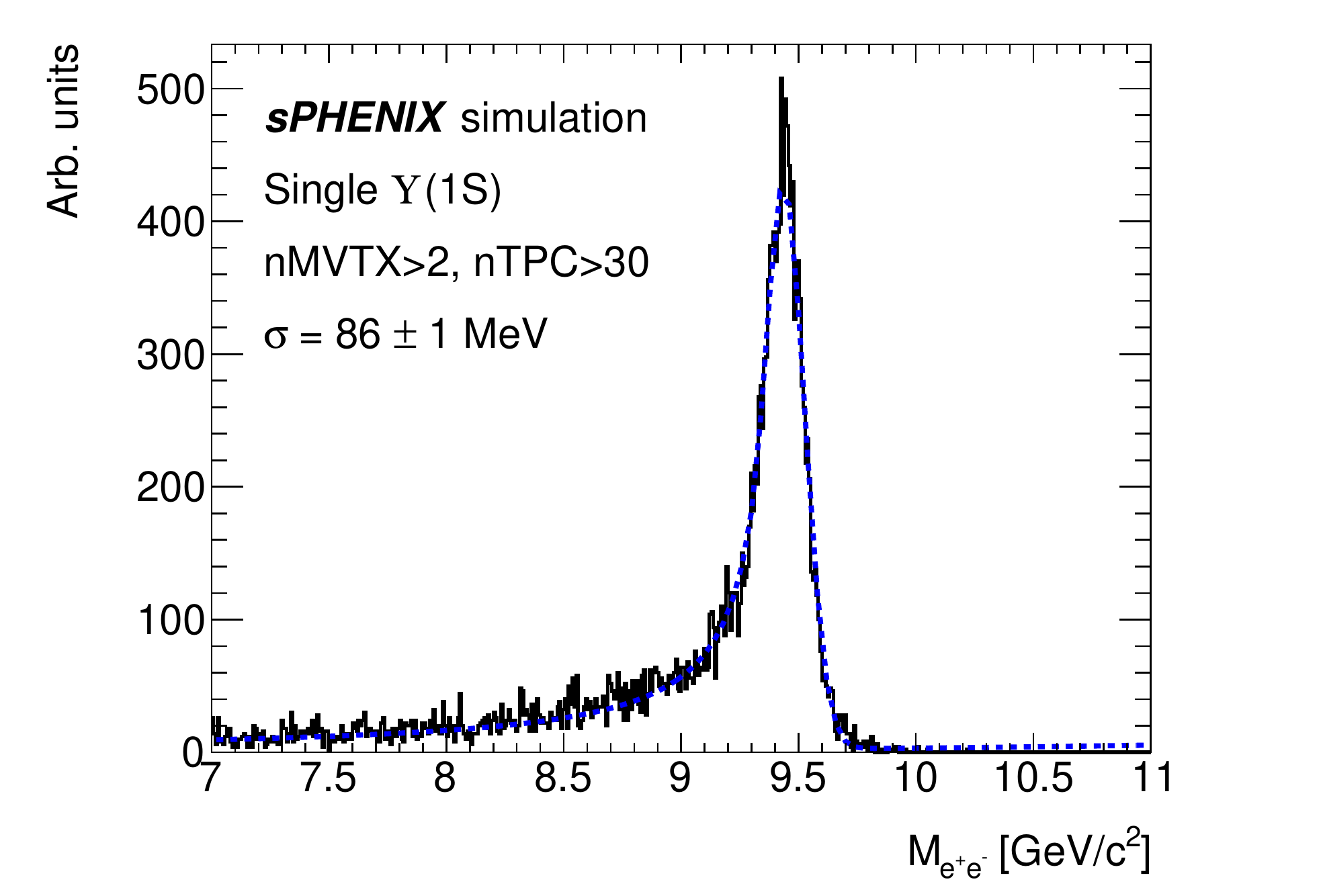}
 	\caption{(Left) The track $p_T$ resolution as a function of $p_T$. (Right) The $\Upsilon(1S)$ mass resolution in single upsilon events. For each figure, the number of MVTX and TPC measurements required per track are shown in the caption.}
 	\label{fig:ptres}
 \end{figure*}

An overview of the track reconstruction workflow can be found in Fig.~\ref{fig:tracking_workflow}. Clusters are found in each subsystem individually, and then used to seed tracklets in the TPC and silicon subsystems. These tracklets are then matched based on azimuthal and pseudorapidity matching windows, such that fully constructed track seeds can be provided to the track fitter. Final tracks are then used to identify the primary collision vertex. The strategy to seed tracks in each subsystem separately and then match them is motivated by the dominant role of the TPC and the large occupancies that sPHENIX will experience. The TPC consists of 48 layers, while the silicon detectors comprise 5 layers. Thus, real tracks that can be found in the TPC contain the vast majority of clusters of a complete sPHENIX track and generally suffer from less combinatoric possibilities. This makes good TPC track seeds a strong foundation for building the track. Because of the small pixel size of the silicon detectors and corresponding excellent cluster resolution, the silicon tracklets are well defined. Thus matching the full silicon tracklet to the full TPC tracklet seems a natural choice for completing the track.

The current sPHENIX track reconstruction strategy uses several ACTS tools in various stages of the reconstruction. First, the ACTS seeding algorithm is run with the measurements in the MVTX. The ACTS seeding algorithm only returns three measurements combined as a ``triplet'' seed, so it is a natural candidate for the MVTX which has three layers. The result of the triplet is propagated to the INTT to find associated measurements to form silicon track seeds.  Figure~\ref{fig:seedEff} shows the ACTS seeding efficiency as implemented in the MVTX for simulated events consisting of 100 pions-per-event thrown in the sPHENIX acceptance, where the efficiency is defined as the fraction of truth tracks with three MVTX hits for which there is at least one seed within the azimuthal and pseudorapidity ranges $\Delta\phi<0.02$ rad and $\Delta\eta<0.006$. The drop in efficiency in the lowest $p_T$ bin is primarily a result of a real loss of efficiency and not simply an artifact of the bin width extending to $p_T=0$ GeV/$c$; however, a small fraction of these tracks are below the effective minimum $p_T$ threshold determined from the sPHENIX magnetic field strength. The seeds are given to the ACTS initial vertex finding algorithm since the silicon layers primarily determine the track position and event vertex resolution in sPHENIX.

 In the TPC, track seeds are found using a cellular automaton seeding algorithm, developed specifically for sPHENIX. This seeding algorithm is based on the algorithm developed for the ALICE TPC\cite{Rohr:2018cxc}. Seeds are found by forming and manipulating a directed graph on the TPC clusters: for each cluster, up to two outgoing edges are assigned, pointing from that cluster to two of its spatially-close neighbors in adjacent layers which together form the straightest triplet. Once this is done for all TPC clusters, the resulting directed graph is pruned such that only mutual edges remain; a mutual edge exists between any pair of clusters that are connected in both directions, such that each cluster has both an ingoing and outgoing edge. This process reduces the directed graph to a collection of individual chains of clusters; each of these cluster chains becomes a candidate track seed. A Kalman filter implementation developed by ALICE \cite{Rohr:2018cxc} provides initial track parameter estimates; based on these initial estimates, the seeds are refined and extended by a track propagation module, developed specifically for sPHENIX. Track seeds containing at least 20 clusters after propagation are selected for further reconstruction. Figure~\ref{fig:caseeder} shows the efficiency of the cellular automaton seeding algorithm, where the efficiency is defined as the fraction of truth tracks for which there is at least one reconstructed seed within the azimuthal and pseudorapidity ranges $\Delta\phi<0.02$ rad and $\Delta\eta<0.006$, respectively.

 These seeds are then connected to the silicon track seeds with azimuthal and pseudorapidity matching criteria. If more than one silicon seed is found to match a TPC seed, the TPC seed is duplicated and a combined full track seed is made for every matched silicon seed. These assembled tracks are provided to the ACTS Kalman Filter track fitting tool. The ACTS fitter takes the full track seed, the estimated track parameters from the seed, and an initial vertex estimate to fit the tracks. Examples of the current track fitting performance are shown in Fig.~\ref{fig:ptres} in simulated events where 100 pions are thrown in the nominal sPHENIX acceptance. The left panel shows the $p_T$ resolution, while the right panel shows the $\Upsilon(1S)$ invariant mass resolution. Both meet the requirements listed in Section 2 in these low multiplicity events. We have found that there is not a significant degradation in physics performance compared to previous sPHENIX track reconstruction implementations. Evaluation of the track reconstruction software in central HIJING~\cite{Gyulassy:1994ew} events with 50 kHz pileup is ongoing. These events represent the highest occupancies that sPHENIX will experience.

\subsection{Track Reconstruction Timing}

\begin{figure}[tbh]
	\centering
	\includegraphics[width=1\linewidth]{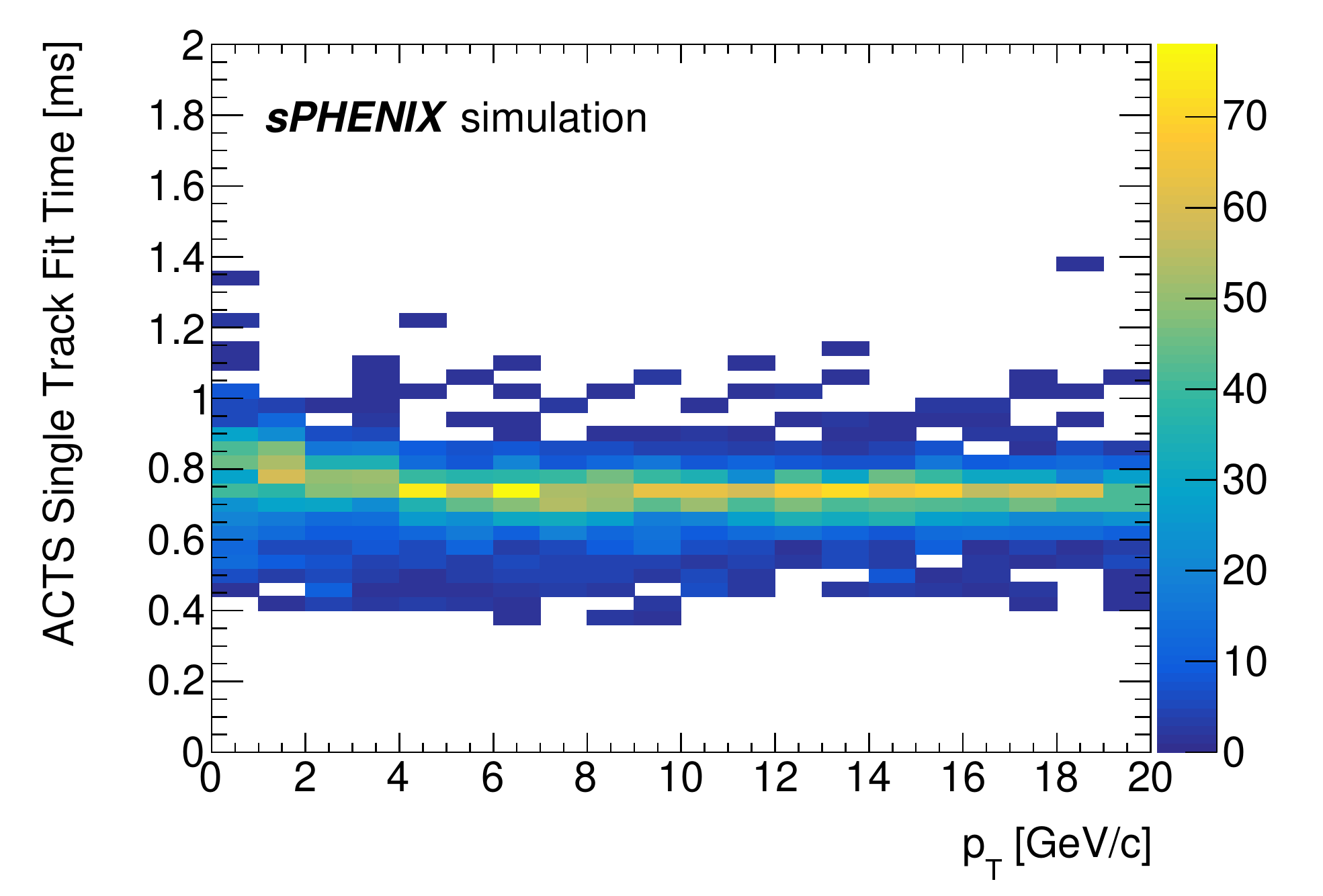}
	\caption{The time per ACTS track fit as a function of $p_T$ in the sPHENIX geometry. This time corresponds to the time taken to run the ACTS track fitting tool on track seeds that were constructed with the various tools described in the text.}
	\label{fig:timetest}
\end{figure}

Another important computational performance test of the ACTS track fitting package is the time spent per track fit. The nominal computational speed goal is to be able to run the track reconstruction in an average of 5 seconds or less per minimum bias event on the BNL computing center that will process the sPHENIX data. Figure~\ref{fig:timetest} shows the time spent per ACTS track fit as a function of $p_T$ for the sPHENIX geometry, which corresponds to approximately 50 layers per track processed by the Kalman filter. The time per track fit is approximately 0.7 ms on average and scales approximately linearly with the number of surfaces the fit visits. For a central Au+Au collision which produces $\sim$1000 tracks, this corresponds to a track fit time of approximately 1 second per event, leaving 80\% of the timing budget for the initial track seeding. At the time of writing, the silicon seeding and TPC track seeding consume approximately equal amounts of time. As track fitting is often one of the more time consuming steps in track reconstruction, this is a major step towards achieving the 5 second total track reconstruction time per event. 

Previous sPHENIX track reconstruction implementations with the GenFit track reconstruction package~\cite{genfit} averaged approximately 80 seconds per minimum bias Au+Au HIJING~\cite{Gyulassy:1994ew} event embedded in 50 kHz pileup. Currently, our track reconstruction implementation with ACTS averages to approximately 10 seconds for the same class of events. One reason for the significant speed up is the way that ACTS handles the material description of the detector. While GenFit uses the full {\sc{Geant 4}} description of the detector to perform material calculations, ACTS uses a condensed and simplified material description that can perform these calculations quickly. 

While the reconstruction timing is a major improvement from the previous sPHENIX track reconstruction implementation, it still does not reach the nominal goal of 5 seconds per minimum bias event. The goal of 5 seconds per event was set as a threshold for processing the data in a timely fashion given the computational resources available at BNL. Since sPHENIX is unique in that it will be a new collider experiment with a nominal run period of 3 years, it is essential that the data be processed as quickly as possible so that time is not wasted on, for example, commissioning the detector. There are several development avenues which will continue to improve this value; for example, our current framework maps an sPHENIX track object to an ACTS track parameters object. Because of this, there is computing time and memory wasted copying the relevant information between the two types. Future development will include switching to an ACTS only data type model, so that the returned ACTS result can be moved directly into storage without copying all of the underlying data types. Another area of expected improvement is in fake track rejection. Currently, time is wasted in attempting to fit fake track seeds which the ACTS track fitter generally rejects as incompatible. Identifying these seeds as fake earlier in the track reconstruction chain will save time wasted in the ACTS Kalman Filter algorithm.

\section{TPC Space Charge Distortions and Alignment}

The sPHENIX TPC measurements will experience significant effects from the build up of positive ions drifting slowly in the TPC towards the central membrane. This space charge will distort the electric fields. Space charge distortions introduce differences between the position measured in the TPC readout detectors and that of the corresponding primary electron that is associated to the true track trajectory. It can be considered as a detector alignment calibration that is specific to TPC geometries. The realism of the track reconstruction procedure presented here is limited by the absence of TPC space charge distortions in the simulation. Implementing space charge distortions in the simulation and correcting for them in the track reconstruction is an ongoing effort within the sPHENIX collaboration. 

Space charge distortion effects will be addressed in several ways. Indirect methods will include continuous monitoring of the charge digitized by the TPC readout electronics, and monitoring the charge collected on the central membrane. More direct methods include using a laser flash to introduce charge of known initial position into the TPC from metal dots on the central membrane, using lasers to ionize the TPC gas in known trajectories, and by measuring the difference between measured TPC cluster positions and tracks extrapolated from external detectors (including the silicon layers).

The latter method will utilize the \texttt{Acts::Propagator} machinery to extrapolate track seeds to the TPC layers from external detectors and estimate their position. The residuals of the actual TPC measurements with the extrapolated position at a given TPC layer will then be determined. This will provide the average distortion in a given $(r,\phi,z)$ bin and time interval. However, the reconstruction of tracks including space charge distortions within ACTS presents a new challenge. Due to the distortion of the measurements, their position on the ACTS surface is inconsistent with the track trajectory. There are two ongoing development strategies to correct the measurement position based on the average distortions. The first is to move the measurements associated with tracks onto the surface, based on the calculated average distortion in a given $(r,\phi,z)$ bin. Since the surfaces are two dimensional, this requires incorporating the radial distortions into the azimuthal and z displacement of the measurement using an estimate of the local track angles. The second is an ACTS three-dimensional fitter which would provide the option for measurements to either be associated to a detector surface or a detector volume. This change in the fundamental philosophy of ACTS measurements is an ongoing development to allow ACTS to handle TPC volumes directly from the geometry implementation. This would allow for measurements to be moved in all three dimensions based on the average distortion. Both strategies are under development and evaluation.

Additionally, the alignment of the detector is an ongoing area of development. This is closely related to the discussion of space charge distortions, since the distortions can be thought of as misaligned measurements in the TPC. Experiment specific adaptations, which can include information like detector conditions and alignment, are made possible through ACTS with \texttt{C++} compile-time specializations. These contextual data types can be specified on an event-by-event basis and are left to individual experiments to define. Properly implementing the alignment calibration techniques, including those of space charge distortions, is one of the challenges that the sPHENIX track reconstruction framework will address in the next year.

\section{Conclusion}\label{conclusion}

The sPHENIX experiment is a high energy nuclear physics experiment being constructed at RHIC to be commissioned in 2022 and begin data taking in 2023. Measuring jets, their substructure, and heavy flavor hadrons in $p$$+$$p$, $p$+Au, and Au+Au collisions are the primary physics observables for the experiment. sPHENIX will collect data in a high rate environment, making track reconstruction with the planned computing resources at Brookhaven National Laboratory a technical challenge. To address these challenges, the track reconstruction software has been completely rewritten to implement various ACTS track reconstruction tools. In this paper, the current performance of this implementation into the sPHENIX software stack has been presented. Due to the constraint that measurements in ACTS must be associated to a surface, it was necessary to add surfaces inside the TPC gas volume, corresponding to the readout layers, that ACTS could associate TPC measurements with. While the configuration of the sPHENIX TPC is not available as a part of ACTS, we note that our implementation is available in Ref.~\cite{sphenixgit} and hope that it serves as a guideline for others attempting to use ACTS with a TPC. Several of the ACTS tools, including seeding, vertexing, and track fitting, are now a part of the default track reconstruction chain. We note that these developments highlight the utility and versatility of the ACTS package, for example by selection of a wide range of tools that are of use for an experiment's chosen track reconstruction strategy. The performance of the track reconstruction in low multiplicity environments has been evaluated and tuned. Tuning of the track seeding and finding in high occupancy events is ongoing. Initial tests with the software described here show an approximately 8x computational speed up from previous sPHENIX track reconstruction software implementations. Additionally, there are several avenues of ongoing development for the handling of space charge distortions in the TPC, both within sPHENIX and by the ACTS developers. These developments will continue to improve the track reconstruction framework as sPHENIX prepares for data taking starting in 2023.


\vspace{12pt}
\noindent\textbf{Acknowledgements}\\
\noindent This work was supported in part by the Office of Nuclear Physics within the U.S. DOE Office of Science under Contract No. DE-SC0012704, DE-SC0013393, and DE-SC0011088, the Commissariat a l’Energie Atomique (CEA), and the National Science Foundation under Award Number 2013115.

\vspace{12pt}
\noindent\textbf{Notice of Copyright}\\
\noindent This manuscript has been authored by UT-Battelle LLC, under contract DE-AC05-00OR22725 with the US Department of Energy (DOE). The US government retains and the publisher, by accepting the article for publication, acknowledges that the US government retains a nonexclusive, paid-up, irrevocable, worldwide license to publish or reproduce the published form of this manuscript, or allow others to do so, for US government purposes. DOE will provide public access to these results of federally sponsored research in accordance with the DOE Public Access Plan (http://energy.gov/downloads/doe-public-access-plan).

\vspace{12pt}
\noindent\textbf{Declarations}\\
\noindent\textit{Conflicts of Interest}~ The authors declare that they have no conflict of interest.\\

\vspace{6pt}

\noindent\textit{Availability of data and material}~ Not  applicable.  No associated data except for code.\\

\vspace{6pt}

\noindent\textit{Code availability}~ The code used for this research is available open source at Ref.~\cite{sphenixgit}.\\
%
%

\bibliographystyle{spphys.bst}       
\bibliography{sPHENIXActsBib}   

\end{document}